\documentclass[letter]{jacow}
\usepackage[utf8]{inputenc}
\usepackage{amsmath}
\usepackage{siunitx}
\usepackage{float}

\ifboolexpr{bool{jacowbiblatex}}
 {
  \addbibresource{jacow-test.bib}
  \addbibresource{biblatex-examples.bib}
 }{}
\title{Improved Measurements of Nonlinear \\Integrable Optics at IOTA}
\author{J. N. Wieland \thanks{wielan22@msu.edu}, A. L. Romanov, A. Valishev. G. Stancari, J. D. Jarvis,  Fermilab, Batavia, IL, USA \\
N. Kuklev, ANL, Lemont, IL, USA \\
S. Szustkowski, LANL, Los Alamos, NM, USA \\
S. Nagaitsev, JLAB, Newport News, VA, USA}
\date{April 2023}

\begin{document}

\maketitle

\begin{abstract}
Nonlinear integrable optics (NIO) are a promising novel approach at improving the stability of high intensity beams. Implementations of NIO based on specialized magnetic elements are being tested at the Integrable Optics Test Accelerator (IOTA) at Fermilab. One method of verifying proper implementation of these solutions is by measuring the analytic invariants predicted by theory. The initial measurements of nonlinear invariants were performed during IOTA run in 2019/20, however the covid-19 pandemic prevented the full-scale experimental program from being completed. Several important improvements were implemented in IOTA for the 2022/23 run, including the operation at higher beam energy of 150 MeV, improved optics control, and chromaticity correction. This report presents on the improved calibrations of the NIO for nonlinear invariant measurements.
\end{abstract}

\section{Introduction}
 Contemporary particle physics experiments benefit from high intensity particle accelerators for primary beams. At high intensities, current accelerators are vulnerable to coherent instabilities. The instabilities may be mitigated by the addition on nonlinear elements for amplitude dependent detuning. Along with general lattice imperfections and higher order effects, these elements serve to reduce the available dynamic aperture. Certain configurations of NIO generate amplitude dependent detuning without impacting the stability of single particle dynamics and reducing the dynamic aperture.
 
 IOTA is a research and development storage ring located at the Fermilab accelerator science and technology facility for studies of practical implementations of NIO. One NIO implementation in IOTA is based on static magnetic elements for a transverse, 4-D integrable system discovered by V. Danilov and S. Nagaitsev (DN) \cite{dn2010}. The theoretical  design follows the T-insert geometry, where where symmetric drift regions with matched horizontal and vertical beta functions are separated by linear sections with an integer multiple of~$\pi$ phase advance. The nonlinear potential is implemented in the former drift of this linear framework. One possible resulting potential was selected for practical implementation and was used to design specialized magnetic elements for IOTA \cite{oshea_measurement_2013,oshea_non-linear_2015,oshea_non-linear_2017}. The theoretical 2-D potential for the DN 
potential \cite{mitchellComplexDN2019} is given by Eq. (\ref{eq:dnPot}), where $t$ is the strength parameter, $c$ is the geometric parameter of the potential, and $\beta(s)$ is the beta function in the T-insert drift.

\begin{equation}
    U(x,y) = t \textrm{Re}\left(\frac{z}{\sqrt{1-z^2}}\textrm{arcsin}(z)\right),\hspace{10pt} z = \frac{x + iy}{c\sqrt{\beta(s)}},
    \label{eq:dnPot}
\end{equation}
 
 There are two stages of the IOTA NIO research program. For the first stage, a low emmittance electron beam produced in the FAST superconducting linac is used to probe the transverse phase space. Parameters of IOTA ring used during the run presented in this paper are listed in the Table~\ref{tab:iota_params}. The second stage will switch to operation with low energy protons.

 \begin{table}
  \centering
  \caption{IOTA Electron Beam Parameters}
    \begin{tabular}{lc}
      \toprule
      Parameter & Nominal Value \\
      \midrule
      Perimeter & 39.96 m  \\
      Energy & 150 MeV  \\
      Betatron tune, ($Q_x, Q_y$) & 5.3 \\
      Equilibrium emittance, ($\epsilon_x,\epsilon_y$) & 30 $nm$ \\
      RF frequency & 30 MHz \\
      Harmonic number & 4 \\
      Synchrotron tune, $\nu_s$ & $3.5 \times 10^{-4}$\\
      Energy spread, $\sigma_E$    &   $1.3\times 10^{-4}$  \\
      Momentum compaction factor & 0.083 \\
      Natural chromaticity ($C_x$, $C_y$) & -10.9, -9.4 \\
      \bottomrule
    \end{tabular}
  \label{tab:iota_params}
\end{table}

 In this paper, the calibrations and verification of the DN nonlinear insert for the current IOTA run are presented.

\section{Bare Lattice Tuning}
\begin{figure}
    \centering
    \includegraphics[width=82.5mm]{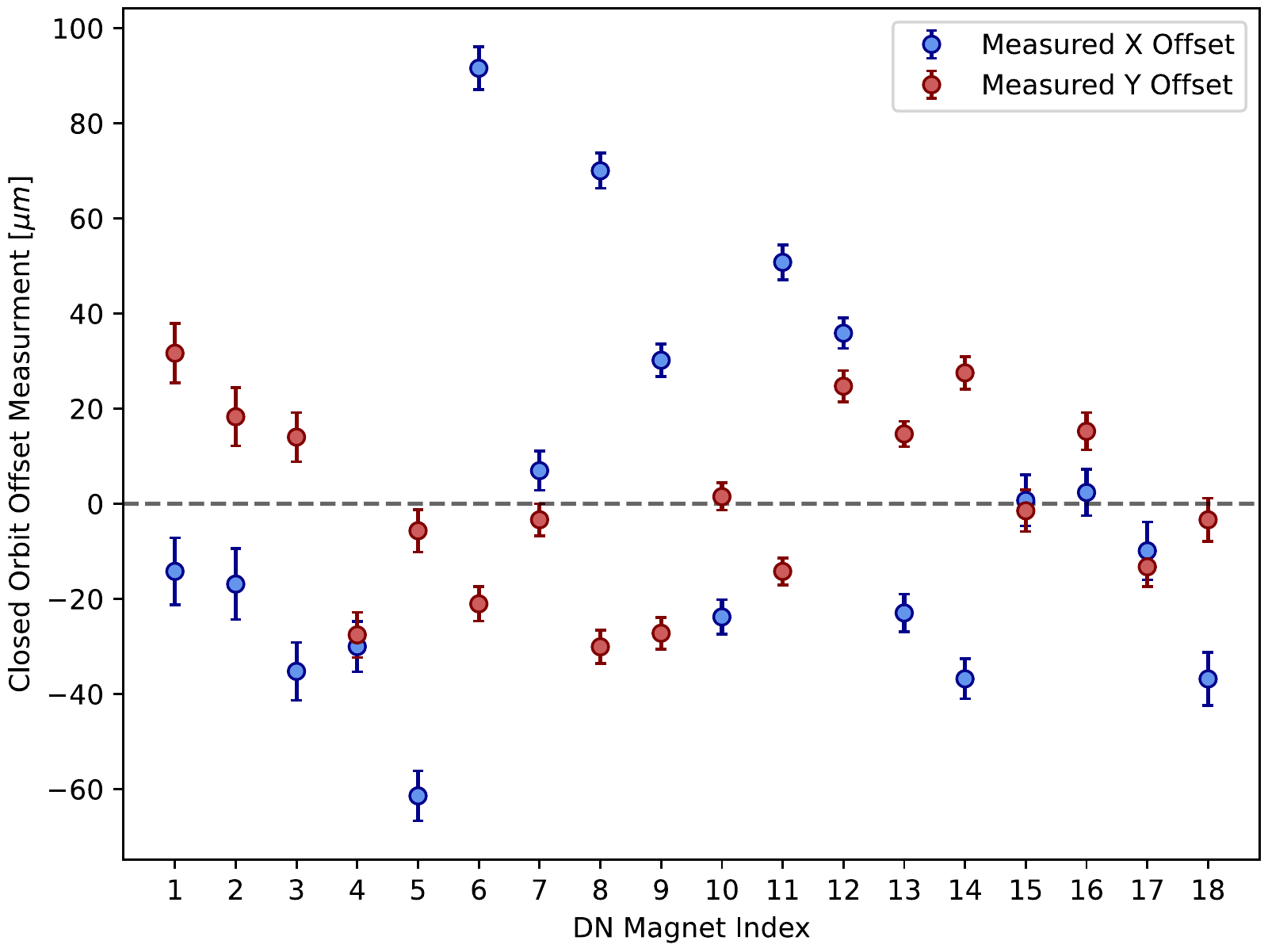}
    \caption{Closed orbit offsets in DN element.}
    \label{fig:dnOffset}
\end{figure}

The requirements of the t-insert geometry place rigorous constraints on the linear lattice configuration. Before taking nonlinear measurements, the bare lattice was extensively tuned using the linear optics from closed orbits (LOCO) algorithm as implemented in Sixdsimulation \cite{romanov_correction_2017}. The measured closed orbit offsets in the DN element are presented in Fig. \ref{fig:dnOffset}.

The measured parameters of interest and their accuracies are included in Table \ref{tab:bareLat}. In addition to linear optics, chromaticity was fully compensated using two families of sextupoles. 

\begin{table}[]
    \centering
    \caption{IOTA Bare Lattice Measurements}   
    \begin{tabular}{lc}
    \toprule
    Bare Lattice Parameter & RMS Accuracy \\
    \midrule
    Betatron Tune & $1\times10^{-5}$  \\
    Insert Phase Advance & $1\times10^{-3}$ \\
    Insert Orbit Centering & $50 \hspace{5pt} $\textmu m \\
    Chromaticity ($C_x,C_y)$ & $0.03,0.06$ \\
    \bottomrule
    \end{tabular}
    \label{tab:bareLat}
\end{table}

\section{DN Element Calibration}
To satisfy the DN potential, the individual elements fields must scale with the beta function. To compensate for variations between the magnets, the 18 individual magnets were calibrated with beam based measurements. The multipole expansion of the nonlinear magnetic field is given in Eq. (\ref{eq:dnMult}). Here $B\rho$ is the beam rigidity, and $t$, $c$, $\beta(s)$ are the same factors as in Eq. (\ref{eq:dnPot}).

\begin{equation}
    B_y + iB_{x} =  -t\frac{B\rho}{\beta(s)} \sum_{n=1}^{\infty}\frac{ 2^{2n-1}n!(n-1)!c}{(2n-1)!\sqrt{\beta(s)}} \left(\frac{x+iy}{c\sqrt{\beta(s)}}\right)^{2n-1}
    \label{eq:dnMult}
\end{equation}

At small amplitudes the quadrupole term dominates. This was verified with measurements of tune dependence on kick amplitude for a single excited magnet. To determine the individual calibration, the quadrupole tune shift due to the first order term is set equal to the measured value and solved for necessary t-scaling, Eq. (\ref{eq:tuneInt}). 

\begin{equation}
    \Delta Q_{x,y} = \pm \frac{1}{4\pi}\int{\beta(s) \frac{\Delta B_2}{B\rho} ds},\hspace{10pt} \Delta B_2 = \frac{-2B\rho \Delta t}{\beta^2(s)}
    \label{eq:tuneInt}
\end{equation}

\begin{figure}
    \centering
    \includegraphics[width=82.5mm]{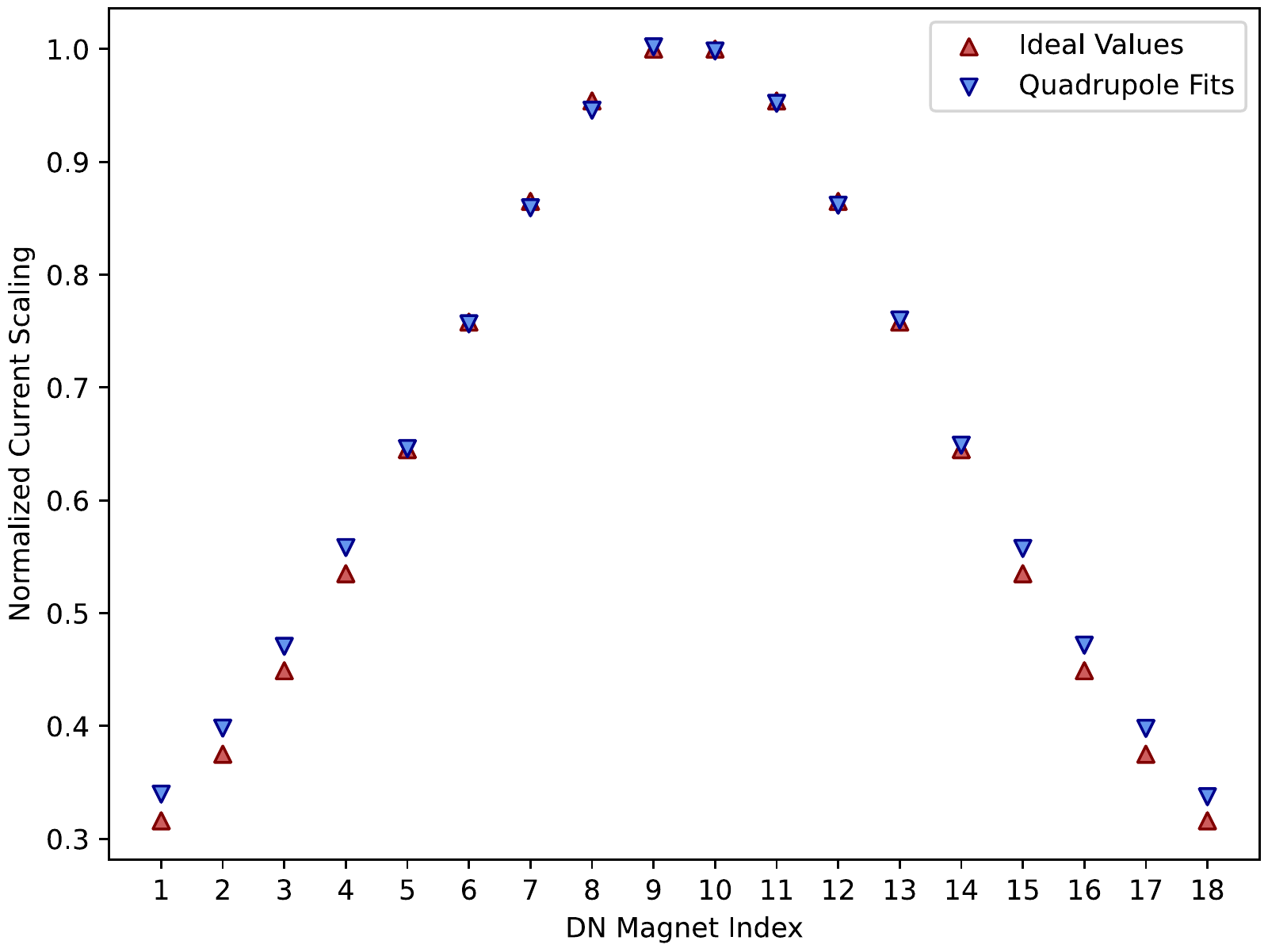}
    \caption{Normalized DN insert current scaling profile compared to calculated ideal values.}
    \label{fig:dnIscaling}
\end{figure}

Each magnet was energized individually and a small amplitude kick was applied to the beam to measure the tune shift. This was done for multiple current setpoints to fit a tune shift vs current. Tune was measured by applying the NAFF algorithm to kicked turn-by-turn bpm data \cite{zisopoulosTune2019}. The resulting data gives a current scaling profile for the overall insert Fig. \ref{fig:dnIscaling}. 

\section{Nonlinear Tune Shift}
The DN insert nonlinear detuning was measured to verify calibration. The theoretical detuning is given in Eq. (\ref{eq:dnDetune}) \cite{nagaitsev_nonlinear_2012}.

\begin{equation}
    \begin{split}
        Q_{x} = Q_{o}\sqrt{1+2t} \\
        Q_{y} = Q_{o}\sqrt{1-2t}
    \end{split}
    \label{eq:dnDetune}
\end{equation}

\begin{figure}
    \centering
    \includegraphics[width=82.5mm]{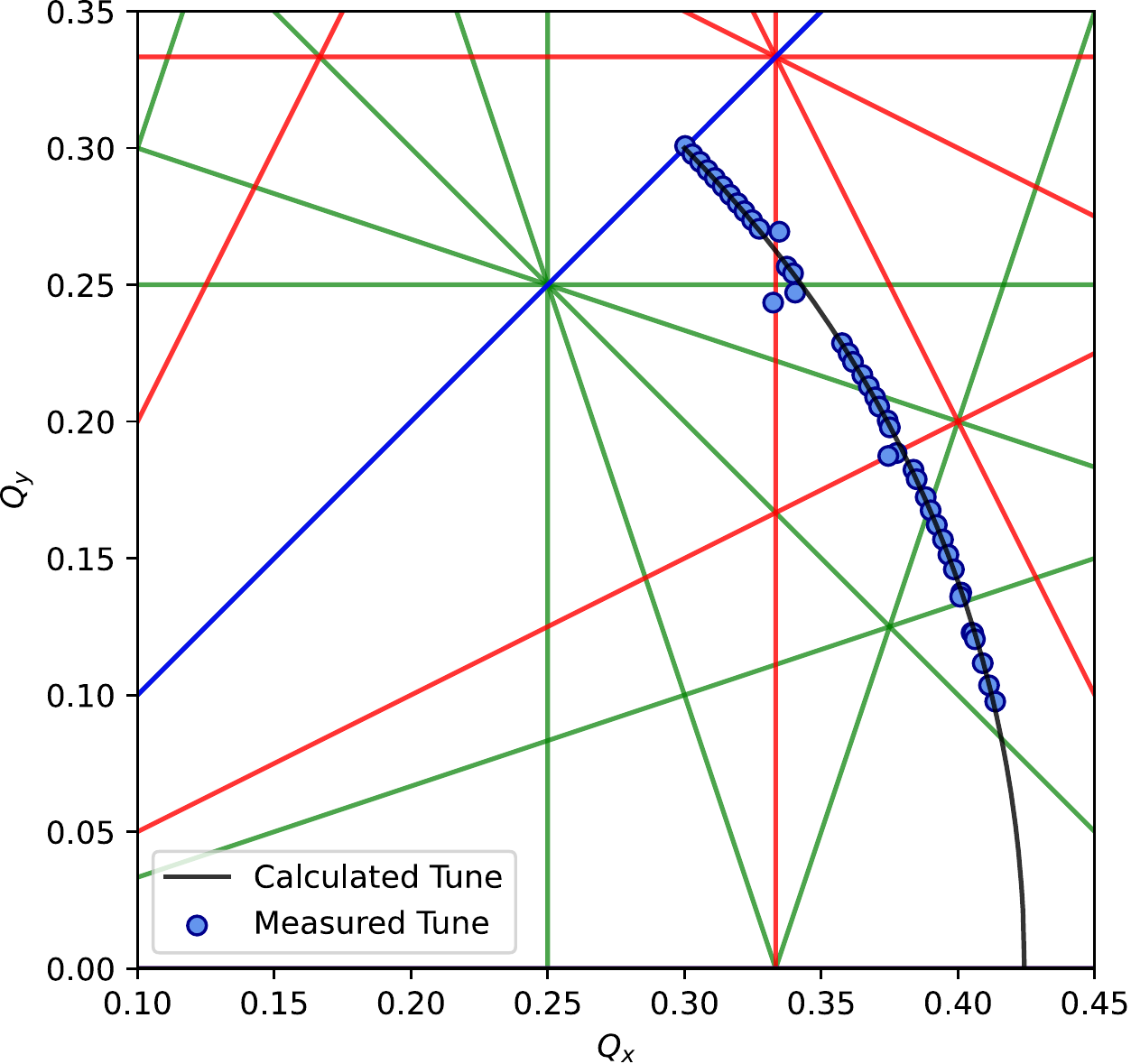}
    \caption{Nonlinear detuning on resonance diagram.}
    \label{fig:dnDetuning}
\end{figure}

The detuning was measured by adjusting the t-parameter of the entire DN insert and kicking the beam for a turn-by-turn tune measurement. The results plotted along with the calculated ideal detuning are presented in Fig. \ref{fig:dnDetuning}.

\begin{figure}
    \centering
    \includegraphics[width=82.5mm]{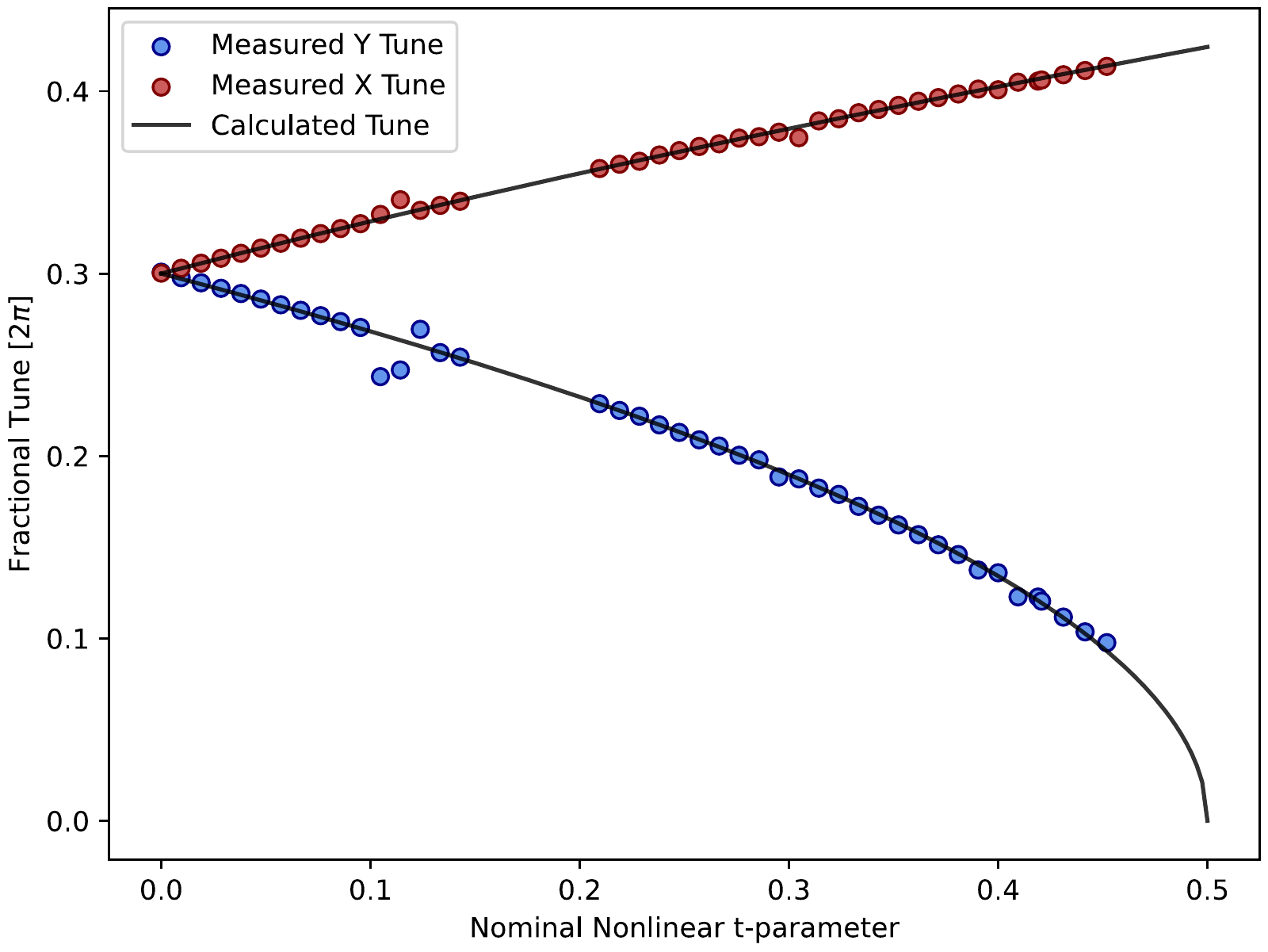}
    \caption{Detuning vs t-parameter in both planes, gap in data between t=0.1 and 0.2 was due to error in data acquisition system, t-parameter of measured data is scaled by 0.935 from fit.}
    \label{fig:dnTuneVsT}
\end{figure}

The detuning ratio between the planes supports proper implementation of the potenetial. The absolute t-parameter scaling may be verified by investigating detuning vs nominal t-parameter, Fig. \ref{fig:dnTuneVsT}. 

The initial data indicated a discrepancy in the tune dependence of the t-parameter. The proportionality between the horizontal and vertical tunes (Fig. \ref{fig:dnDetuning}) showed no significant aberrations, which pointed to the absolute scaling of the t-parameter. The ideal detuning expression was fit to the data with a t-scaling factor as the only free parameter Eq. (\ref{eq:fitTune}). 

\begin{equation}
    Q = Q_o \sqrt{1\pm 2at}
    \label{eq:fitTune}
\end{equation}

This indicated that the calibrated t-parameter scaling was $a \approx 0.935$ of the nominal t-parameter. The source of this discrepancy is not yet clear, but is suspected to be a combined effect of the entire insert since the calibration was measured with single elements.

The impact of the strong sextupoles implemented to compensate chromaticity can be seen in Fig. \ref{fig:dnDetuning} as the dominant effect on the tune near the third order resonances (red lines on tune diagram).

\section{Losses Scan}
\begin{figure}
    \centering
    \includegraphics[width=82.5mm]{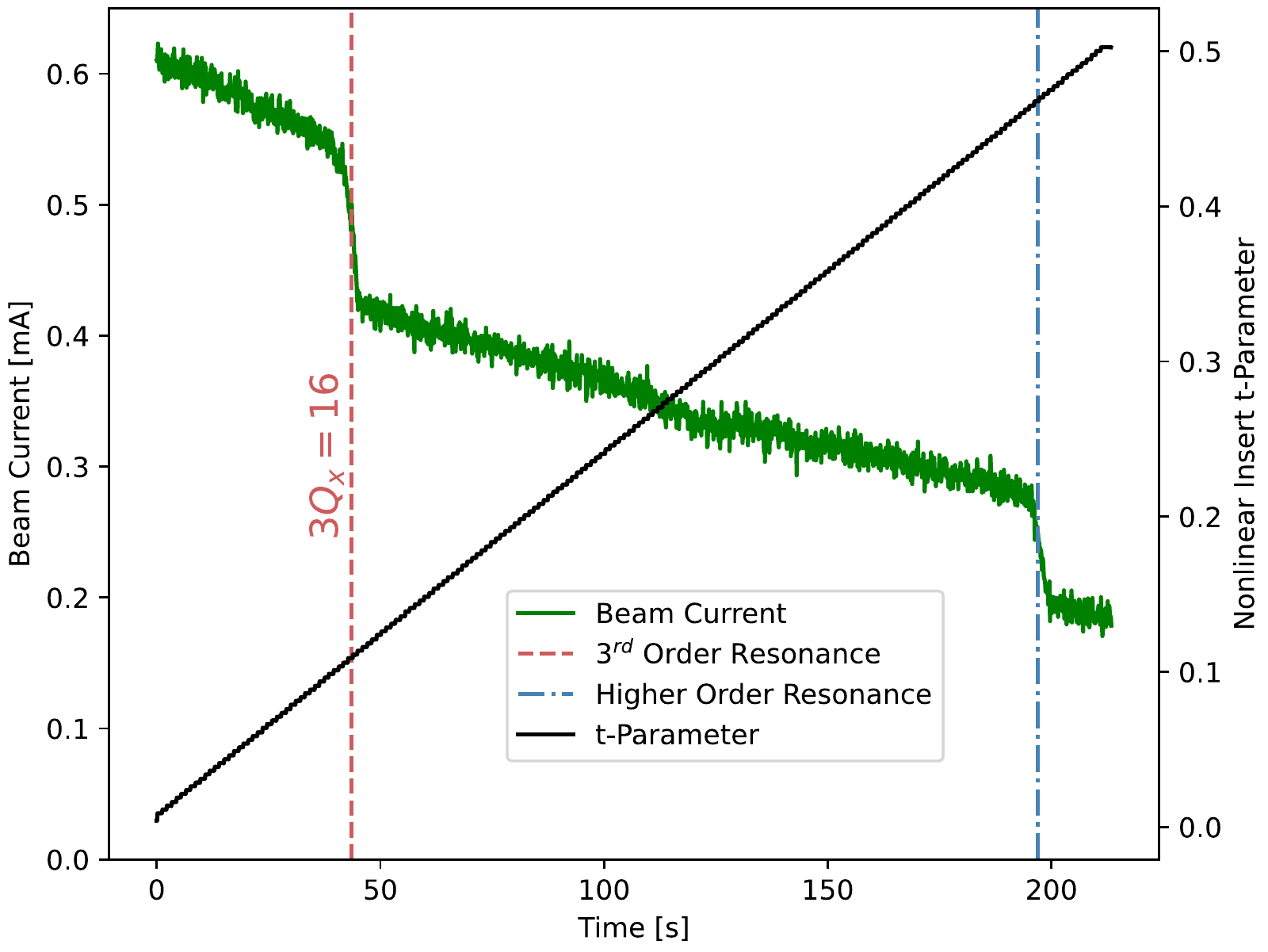}
    \caption{Beam current for t-parameter scan.}
\label{fig:iVsT}
\end{figure}

A final verification of the implementation was performing a scan of the t-parameter and logging the beam current to determine losses, Fig. \ref{fig:iVsT}. 

The largest losses can be seen at the t-parameters corresponding to crossing the horizontal third order resonance. The third order resonances show some improvement over the 2019/2020 run where a sextupole scaling scheme had to be implemented in order to cross these resonances without ~90\% beam loss if they were energized. In practical operation, this is not an issues as losses are negligible if the resonance is crossed quickly. Additional losses are observed at a t-parameter of ~0.46. There are a several higher order resonances which may drive losses at this point but the exact source is not clear. There are no significant losses observed at the integer resonance.

\section{Conclusion}
The bare lattice and DN NIO insert in IOTA have been calibrated and verified for studies. Data collection for calculation of the nonlinear invariants from reconstructed turn-by-turn phase space is now ongoing.

%\bibliography{WEPL052}

\begin{thebibliography}{1}
	
	\bibitem{dn2010}
	V.~Danilov and S.~Nagaitsev, ``Nonlinear accelerator lattices with one and two
	analytic invariants,'' {\em Phys. Rev. Spec. Top. Accel Beams},  vol.~~3, p.~084002, 2010. \\
	\url{doi:10.1103/PhysRevSTAB.13.084002}
	
	\bibitem{oshea_measurement_2013}
	F.~H. O'Shea, R.~B. Agustsson, A.~Y. Murokh, and E.~Spranza, \textquotedblleft{Measurement of Non-Linear Insert Magnets}\textquotedblright,
	in \emph{Proc. NAPAC’13}, Pasadena, CA, USA, Sep.-Oct. 2013, paper WEPBA17, pp. 922--924. 
	
	\bibitem{oshea_non-linear_2015}
	F.~O'Shea, R.~Agustsson, Y.-C. Chen, D.~Martin, J.~McNevin, and E.~Spranza,
	\textquotedblleft{Non-linear Magnetic Inserts for the Integrable Optics Test Accelerator}\textquotedblright,
	in \emph{Proc. IPAC’15}, Richmond, VA, USA, May 2015, pp. 724--727. \\
	\url{doi:10.18429/JACoW-IPAC2015-MOPMN010} 
	
	\bibitem{oshea_non-linear_2017}
	F.~H. O'Shea, R.~B. Agustsson, P.~S. Chang, and Y.~C. Chen, \textquotedblleft{Non-Linear Inserts for the IOTA Ring}\textquotedblright,
	in \emph{Proc. IPAC’17}, Copenhagen, Denmark, May 2017, pp. 4407--4409. \\
	\url{doi:10.18429/JACoW-IPAC2017-THPIK129} 
	
	\bibitem{mitchellComplexDN2019}
	C.~Mitchell, ``Complex representation of potentials and fields for the
	nonlinear magnetic insert of the integrable optics test accelerator,'' {\em
		arXiv}, 2019. \\ \url{doi:10.48550/arXiv.1908.00036}
	
	\bibitem{romanov_correction_2017}
	A.~Romanov, D.~Edstrom~Jr., F.~A. Emanov, I.~A. Koop, E.~A. Perevedentsev,
	Y.~A. Rogovsky, D.~B. Shwartz, and A.~Valishev, ``Correction of magnetic
	optics and beam trajectory using {LOCO} based algorithm with expanded
	experimental data sets,'' {\em arXiv}, Mar. 2017. \url{doi:10.48550/arXiv.1703.09757}

	
	\bibitem{zisopoulosTune2019}
	P.~Zisopolous, Y.~Papahilippou, and J.~Laskar, ``Refined betatron tune
	measurements by micking beam position data,'' {\em Phys. Rev. Spec. Top. Accel Beams}, vol. 22, p. 071002, 2019. \url{doi:10.1103/PhysRevAccelBeams.22.071002}
	
	\bibitem{nagaitsev_nonlinear_2012}
	S.~Nagaitsev, A.~Valishev, and V.~Danilov, ``Nonlinear optics as a path to
	high-intensity circular machines,'' {\em arXiv}, July 2012.
 \url{doi:10.48550/arXiv.1207.5529}
	
\end{thebibliography}
%\bibliographystyle{ieeetr}

\end{document}